\documentclass[aps,prl,twocolumn,groupedaddress,showpacs]{revtex4}

\usepackage{graphicx}
\usepackage{dcolumn}
\usepackage{bm}

\begin{document}

\title{Evanescent-wave trapping and evaporative cooling of an atomic gas\\near two-dimensionality}

\author{M. Hammes}
\author{D. Rychtarik}
\author{B. Engeser}
\author{H.-C. N\"agerl}
\author{R. Grimm}

\affiliation{Institut f\"ur Experimentalphysik, Universit\"at
Innsbruck, Technikerstr.\ 25, A-6020 Innsbruck, Austria}

\date{August 16, 2002}

\begin{abstract}
A dense gas of cesium atoms at the crossover to two-dimensionality
is prepared in a highly anisotropic surface trap that is realized
with two evanescent light waves. Temperatures as low as 100\,nK
are reached with 20.000 atoms at a phase-space density close to
0.1. The lowest quantum state in the tightly confined direction is
populated by more than 60\%. The system offers intriguing
prospects for future experiments on degenerate quantum gases in
two dimensions.

\end{abstract}


\maketitle

Systems with reduced dimensionality have attracted considerable
attention in various areas of physics as they exhibit strikingly
different properties in comparison to the three-dimensional case.
In the field of atomic quantum gases, there has been long-standing
interest in two-dimensional systems in view of the phase
transition to a quantum-degenerate gas \cite{Bagnato1991a}, the
onset of coherence \cite{Kagan1987a}, and modifications of
interaction properties \cite{Petrov2000a}. Spin-polarized hydrogen
on a liquid-helium surface represents a 2D quantum gas as the
adsorbed atoms occupy a single quantum state \cite{Walraven1991a},
and evidence of quantum degeneracy has been obtained
\cite{Safonov1998a}. The required cryogenic set-up and the lack of
efficient optical diagnostics, however, makes detailed
experimental studies on 2D gases of hydrogen very difficult, and
thus there is great interest to realize analogous systems by using
more readily accessible laser-manipulated atoms.

Few experiments with laser-cooled atoms have so far approached 2D.
In standing-wave traps conditions were obtained with the
predominant population of the lowest quantum state in one
dimension \cite{Vuletic1998a, Bouchoule1999a, Bouchoule2002a}. In
such traps, however, many individual traps are populated
simultaneously so that phenomena related to individual systems are
hardly accessible. Another approach is an anisotropic optical trap
made of an elliptically focused laser beam and loaded with a
Bose-Einstein condensate \cite{Gorlitz2001a}; this trap however is
diffraction-limited and thus provides relatively small level
spacing. Optical surface traps combine tight confinement in one
direction with a single potential well. Recent surface trapping
experiments \cite{Gauck1998a, Hammes2001a} have approached 2D by
populating a few bound states, yet far away from quantum
degeneracy.

In this Letter, we report on trapping and evaporative cooling of a
dense gas of Cs atoms in the combined field of two evanescent
light waves.
In this highly anisotropic surface trap, evaporative cooling is
performed by ramping down the trap potential, and temperatures as
low as 100\,nK are reached with 20.000 atoms at a phase-space
density close to 0.1. In the tightly confined direction a
ground-state population of more than 60\% is achieved, and a gas
at the crossover to two-dimensionality is realized.

\begin{figure}
\includegraphics[width=7.2cm]{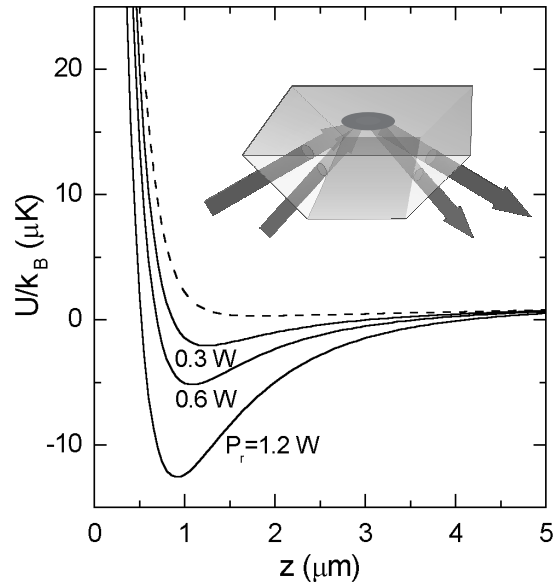}
\caption{\label{trap} Illustration of the DEW trap and calculated
potentials normal to the surface in the trap center. The
attractive van-der-Waals part at short distances
($z\lesssim100\,$nm) is not shown. The solid lines refer to
different values of the power $P_{\rm r}$ of the laser beam
generating the attractive EW.
The dashed line shows the potential for $P_{\rm r} = 0$.}
\end{figure}

The double evanescent-wave trap (DEW trap, see illustration in
Fig.~\ref{trap}) was suggested by Ovchinnikov, Shul'ga, and
Balykin already in 1991 \cite{Ovchinnikov1991a} and has been
discussed in various contexts, like  
2D gases \cite{Desbiolles1996a, Lorent2002a}, atom coupling to
microspheres \cite{Mabuchi1994a}, cavity-induced cooling
\cite{Domokos2001a}, quantum reflection \cite{cote2002a},
integrated optical waveguides \cite{Barnett2000b}, and nano-traps
\cite{Burke2002a}. Despite of this considerable interest, the DEW
trap has not been experimentally realized until now, which may be
explained by the difficulty of loading a narrow potential well
very close to a material wall. We have accomplished efficient
loading by transfer from a cold and dense reservoir of cesium
atoms. In our case, this is provided by the non-dissipative
optical surface trap introduced in Ref.~\cite{Hammes2002a}. The
demonstrated principle of loading from a dense reservoir is
universal and may be also implemented with other optical or
magnetic trapping schemes.

The DEW trap relies on the optical dipole force in a combination
of a repulsive and an attractive EW field, produced on the surface
of a dielectric prism by total internal reflection of two laser
beams. The two fields with wavelengths $\lambda_i$ $(i={\rm
r,\,b})$ are created by laser sources tuned below and above the
atomic transition (red and blue detuning), respectively. Normal to
the surface the two fields decay exponentially with characteristic
lengths of $\Lambda_i = \lambda_i/2\pi \times
(n^2\sin^2\theta_i-1)^{-1/2}$, determined by the angle of
incidence $\theta_i$, the optical wavelength $\lambda_i$, and the
refractive index $n$. A narrow, wavelength-sized potential well is
created close to the surface when the decay length $\Lambda_{\rm
b}$ of the repulsive EW is short compared to $\Lambda_{\rm r}$ of
the attractive field. This is reached by setting the red-detuned
laser beam near to the critical angle $\theta_{\rm c} =
\arcsin(1/n)$ of total internal reflection and the blue one much
further away. Typical angles are $\theta_{\rm r} - \theta_{\rm c}$
of a few tens of a degree and $\theta_{\rm b} - \theta_{\rm c}$ of
a few degrees. In addition to this tight confinement normal to the
surface, lateral confinement is achieved when the extension of the
attractive EW field is smaller than the one of the repulsive EW
field; this can be obtained by an appropriate choice of the
Gaussian beam waists $w_i$ of the two applied laser beams.

In our experiment, the repulsive EW is generated with the beam of
a Titanium-Sapphire laser at a wavelength of $\lambda_{\rm b} =
850.5\,$nm (1.6nm blue detuning with respect to the Cs D$_2$ line
at 852.1\,nm). The angle of incidence is set to $\theta_{\rm b} =
46.8^{\circ}$, i.e.\ $3.2^{\circ}$ above the critical angle of the
fused-silica prism ($n=1.45$). The beam for the attractive EW is
derived from an Yb fiber laser with $\lambda_{\rm r} = 1064\,$nm
and set to 0.2$^{\circ}$ above $\theta_{\rm c}$. The resulting
decay lengths are $\Lambda_{\rm b} = 395$\,nm and $\Lambda_{\rm r}
= 2.0\,\mu$m. The beams are focused to waists of $w_{\rm b} =
400\mu$m and $w_{\rm r} = 160\mu$m.

The total trap potential can be written as
\begin{eqnarray}
\label{potential}
U(\mathbf r) & = & U_{\rm b}(x,y) {\rm e}^{-2z/\Lambda_{\rm b}}
- U_{\rm r}(x,y) {\rm e}^{-2z/\Lambda_{\rm r}}\\
\nonumber & & - \alpha_{\rm vdW} z^{-3}(1+2 \pi z/\lambda_{\rm
eff})^{-1} + mgz \nonumber
\end{eqnarray}
where $U_i(x,y)=\hat{U}_i \exp(-2x^2/w_i^2 - 2y^2/w_i^2
\cos^2\theta_i$) are the optical potentials directly at the
surface ($z=0$). For the maximum potentials in the center of the
trap we calculate $\hat{U}_{\rm b}/k_B=325\,\mu$K and
$\hat{U}_{\rm r}/k_B=43\,\mu$K at typical beam powers of $P_{\rm
b} = 1.15\,$W and $P_{\rm r} = 1.2\,$W, respectively; $k_B$
denotes Boltzmann's constant. The laser beams are linearly
polarized in the plane of incidence (TM polarization) to maximize
the EW field. The third term in Eq.~\ref{potential} describes the
van der Waals surface attraction \cite{Landragin1996a,
Shimizu2001a} with a coefficient $\alpha_{\rm vdW} =
2.8\times10^{-49}$\,kg\,m$^5$\,s$^{-2}$ and an effective
wavelength $\lambda_{\rm eff}= 866\,$nm corresponding to the line
center of the D-line dublett; this surface attraction is found to
play a minor role for our experimental conditions. The last term
in Eq.~\ref{potential} represents the gravitational potential with
$m$ denoting the mass of a Cs atom and $g$ representing the
gravitational acceleration.

Fig.~\ref{trap} shows the calculated potential in the center of
the trap ($x,y=0$). 
For $P_{\rm r} = 1.2\,$W, the power that we use for loading, a
13\,$\mu$K deep well is realized with its minimum located
0.9$\mu$m away from the surface. For decreasing power of the
attractive EW the well becomes shallower and the minimum moves
away from the surface. Without attractive EW (dashed line) the
vertical motion is still confined because of gravity, but without
any horizontal confinement.

The multi-stage loading sequence proceeds as follows: Atoms are
released from a magneto-optical trap (MOT) into the
gravito-optical surface trap (GOST) of
Ref.~\cite{Ovchinnikov1997a}, where an unpolarized sample in the
lower hyperfine state ($F=3$) is produced by evanescent-wave
Sisyphus cooling. A part of the atoms is then transfered through
elastic collisions into a narrow, far-detuned intense laser beam
perpendicularly intersecting the evanescent-wave in the center of
the GOST \cite{Hammes2002a}; this beam has a Gaussian beam waist
of 160$\mu$m and is derived with an initial power of 7.2\,W from
the same Yb-fiber laser that later on produces the attractive
field of the DEW trap. The repulsive evanescent wave applied at
this stage is already the one that is later used in the DEW trap.
The sample is then further cooled adiabatically and evaporatively
by exponentially ramping down the power of the horizontally
confining beam to 2\,W in 2\,s. In this way, we obtain our surface
reservoir of $1.8 \times 10^6$ atoms at a temperature of
3.0\,$\mu$K, corresponding to a peak number density of
$\sim$$10^{13}$\,cm$^{-3}$. The peak phase-space density is
$\sim$$10^{-3}$ under the assumption of a fully unpolarized sample
equally distributed among the seven magnetic sublevels.


Transfer into the DEW trap from the reservoir is then accomplished
in a time interval of 50\,ms by
ramping down the power of the red-detuned beam from 2\,W to zero
simultaneously with ramping up the power of the attractive EW from
zero to the optimum loading power $P_{\rm r} = 1.2$\,W. The
lateral confinement of the reservoir beam and the DEW trap are
approximately matched, so that the loading process is governed by
the vertical motion. Up to $10^5$ atoms are observed in the DEW
trap after 150\,ms when the unbound atoms remaining from the
reservoir have laterally escaped. Measurements of the atom number
are performed by recapture into the MOT and detection of the
integrated fluorescence signal by means of a calibrated CCD
camera; the systematic calibration error is below a factor of 1.5.
The observed transfer ratio of $\sim$5\% roughly corresponds to
the volume ratio of the DEW trap and the reservoir.

\begin{figure}
\includegraphics[width=8.5cm]{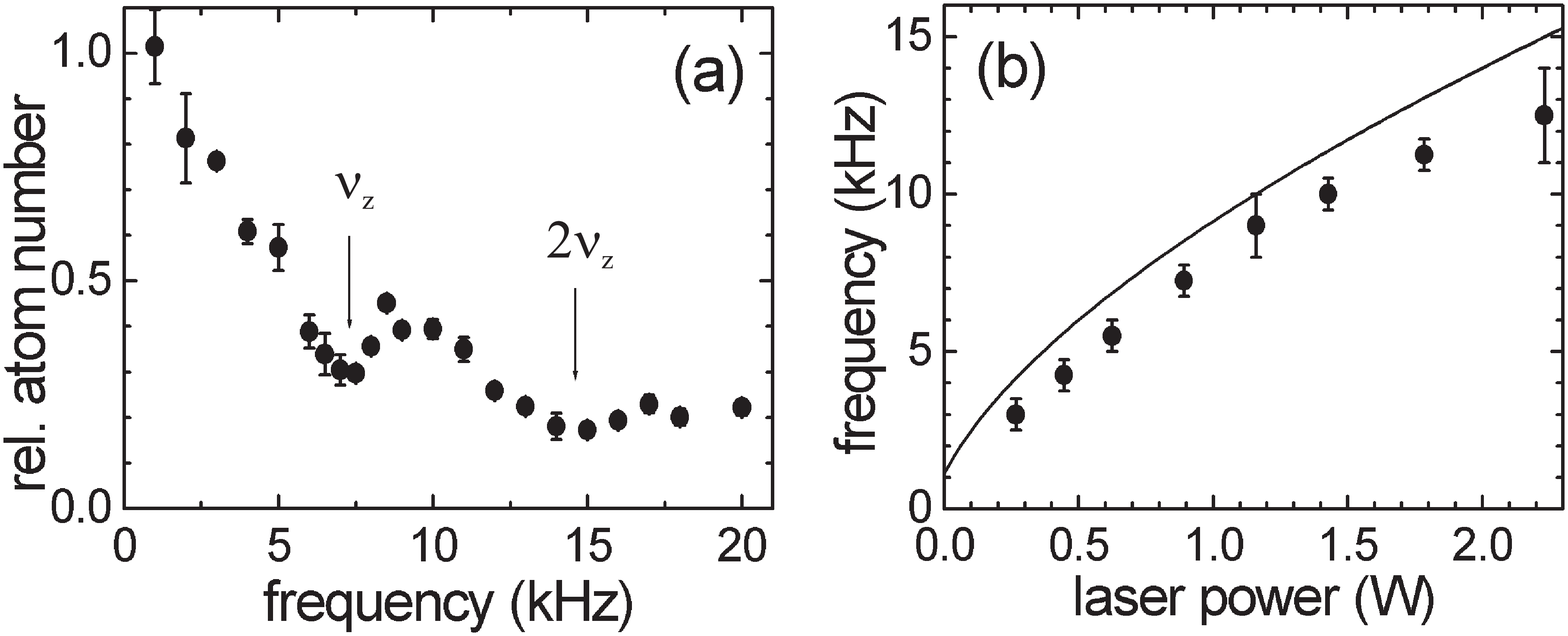}
\caption{\label{trapfrequency} Measurements of the trap frequency
in the tightly confined $z$-direction. (a) Fraction of atoms
remaining trapped for $P_{\rm r} = 0.9\,$W after 150\,ms of
parametric excitation at a variable frequency. (b) Measured trap
frequency versus $P_{\rm r}$ in comparison with a calculation
based on Eq.~\ref{potential}.}
\end{figure}

A very important parameter is the vertical trap frequency
$\nu_{z}$, corresponding to the vibrational energy quantum $h
\nu_{z}$ in the tightly confined direction. We measure the trap
frequency as a function of the power $P_{\rm r}$ by parametric
heating \cite{Friebel1998a}. After loading the DEW trap under
fixed conditions ($P_{\rm r}= 1.2\,$W), we ramp $P_{\rm r}$ to a
variable value in 50\,ms and apply a sinusoidal power modulation
with a typical depth of 10\% to the repulsive EW for 150\,ms. We
then wait 100\,ms to let unbound atoms escape and measure the
number of remaining atoms. A typical trap-frequency measurement is
shown in Fig.~\ref{trapfrequency}(a) for $P_{\rm r} = 0.9\,$W, it
exhibits clear minima at the trap frequency and its second
harmonics. Fig.~\ref{trapfrequency}(b) shows the measured values
for the trap frequency in comparison with the results of a
calculation according to Eq.~\ref{potential}; the latter is based
on a harmonic approximation to the minimum of the trap potential
$U(\mathbf r)$ which we calculate for our experimental conditions
without any adjustable parameter. The measurements are affected by
the lateral thermal spread of the sample and by anharmonicities of
the trap and therefore show an sample-averaged frequency which is
slightly below the frequencies in the trap center. If this minor
deviation is taken into account the measurements well confirm the
theoretical frequencies. At $P_{\rm r} = 1.2$\,W, we calculate a
vertical trap frequency of 10.2\,kHz and horizontal frequencies of
59\,Hz and 42\,Hz. The typical aspect ratio of the trap is thus of
the order of 200.

\begin{figure}
\includegraphics[width=6.5cm]{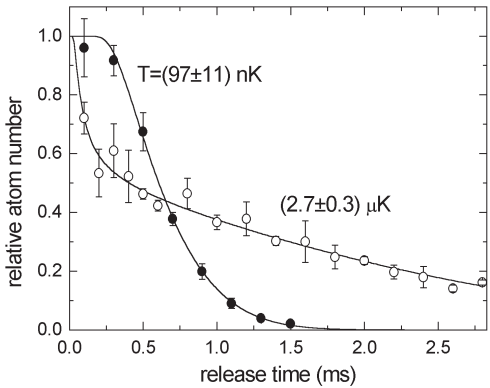}
\caption{\label{temperature} Temperature measurements in the DEW
trap after 300\,ms of storage at constant trap depth ($\circ$) and
after forced evaporative cooling ($\bullet$).}
\end{figure}

The temperature of the sample is measured with a
release-and-recapture method. Both EW fields are turned off
simultaneously by acousto-optical modulators. The sample then
undergoes a ballistic expansion in the field of gravity and atoms
hitting the room-temperature surface are immediately lost. After a
short release time the repulsive EW is turned on again to prevent
further atoms from hitting the surface, and the number of
remaining atoms is measured in the standard way by recapture into
the MOT. A theoretical model
 based on the assumption of a thermal distribution
in the known trap potential is then used to fit the data with a
single parameter and thus to accurately determine the mean kinetic
energy of the released sample. At temperatures down to a few
100\,nK a classical approximation holds and the model directly
provides the temperature. At lower temperatures the quantum nature
of the vertical motion has to be taken into account. We find that,
for all measurements reported here, corrections to the classical
approximation due to the zero-point energy and the discrete level
spacing stay well below 20\% and can be introduced within a simple
harmonic oscillator model. Fig.~\ref{temperature} shows two
examples of temperature measurements obtained for a `hot'
(2.7\,$\mu$K) and a `cold' (100\,nK) sample.



Forced evaporative cooling is implemented by ramping down the
power $P_{\rm r}$ of the attractive EW. In this way, a
two-dimensional evaporation scheme is realized in which energetic
atoms can escape horizontally from the trap. In the vertical
direction no evaporation can take place because of the
gravitational potential. The scheme decompresses the sample and
provides simultaneous adiabatic and evaporative cooling; thus a
drastic reduction of the temperature is realized. Immediately
after loading we apply an exponential ramp that reduces $P_{\rm
r}$ to 3.5\% of its initial value in 400\,ms. As it takes about
150\,ms for the untrapped reservoir atoms to disappear, we can
perform measurements on the number $N$ and temperature $T$ of the
trapped sample for evaporation times $t_{\rm ev} \ge 150\,$ms.

\begin{figure}
\includegraphics[width=7cm]{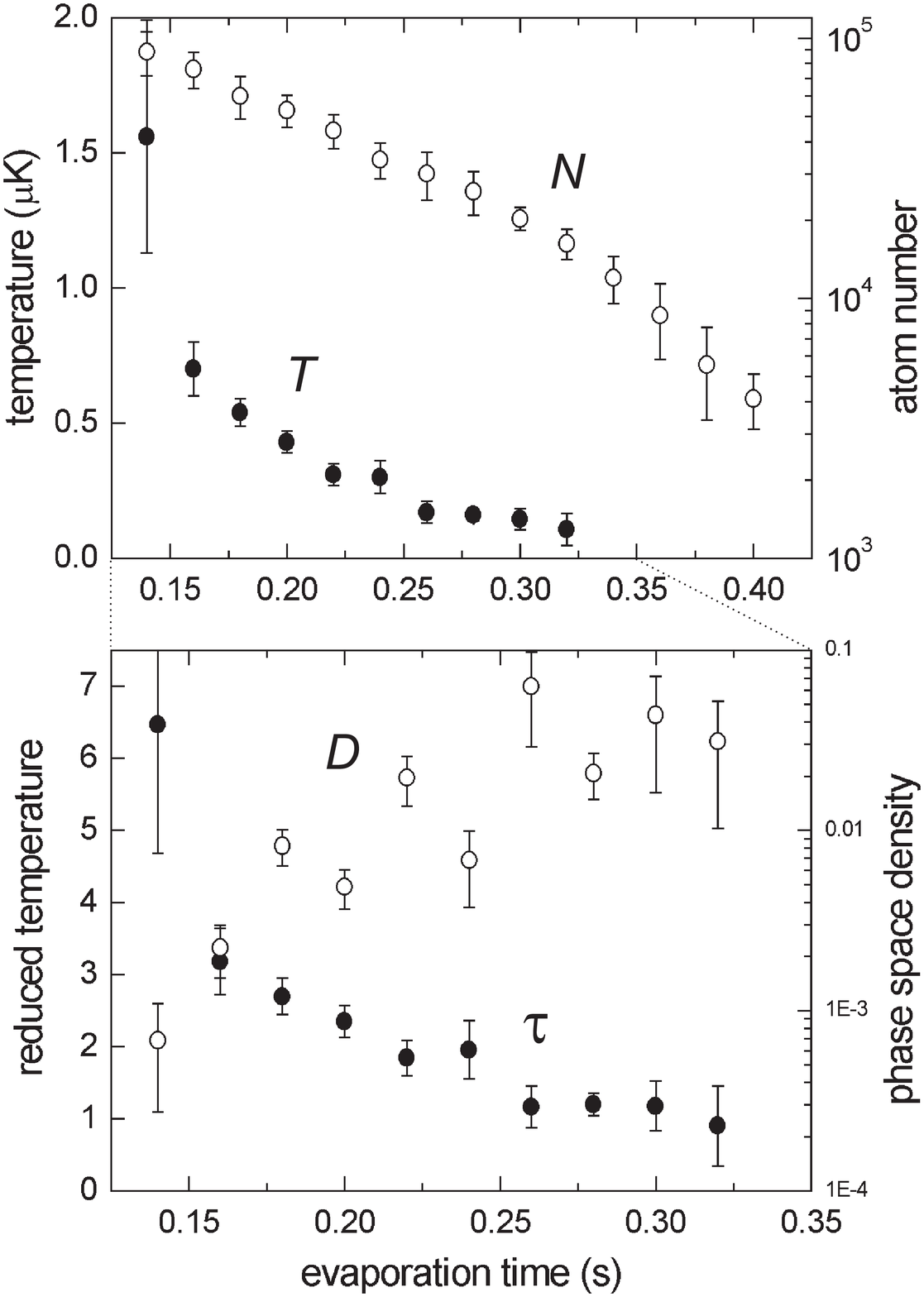}
\caption{\label{evaporation}Results of evaporative cooling in the
DEW trap. Upper graph, temperature $T$ and atom number $N$ versus
evaporation time. Lower graph, reduced temperature
$\tau=k_BT/h\nu_z$ and phase-space density $D$.}
\end{figure}

Fig.~\ref{evaporation} displays the forced evaporation results.
For $t_{\rm ev}$ between 150\,ms ($P_{\rm r}$ reduced to 28\%) and
300\,ms (8\%), an exponential decay of the trapped atom number $N$
is observed. 
In this time interval, the temperature drops from $\sim$1$\mu$K to
$\sim$100\,nK. For 150\,ms $\le t_{\rm ev}\le$ 270\,ms the
phase-space density $D$ of the sample increases by two orders of
magnitude from $\sim$$10^{-3}$ to $\sim$0.1 with a reduction of
$N$ by only a factor of three. This indicates a highly efficient
evaporation process at this stage. At $t_{\rm ev} \approx 270\,$ms
the temperature data level off at $T \approx 100\,$nK, and for $
t_{\rm ev} \ge 300$\,ms an increased loss is observed and reliable
temperature measurements are no longer possible. The highest
attained phase-space densities are of the order of 0.1, when an
equal distribution among the seven magnetic sublevels is assumed.
This compares to the highest phase-space densities so far obtained
with cesium \cite{Han2001a}.

In view of a 2D gas, we discuss the results in terms of a reduced
temperature $\tau = k_B T / h \nu_z$ by relating the thermal
energy to the vibrational energy quantum of the tightly confined
direction. For this purpose, we use the calculated $\nu_z(P_{\rm
r})$ as displayed in Fig.~\ref{trapfrequency}(b). The lower graph
of Fig.~\ref{evaporation} shows that we are presently able to
reach the condition $\tau = 1$. This already corresponds to a mean
vibrational quantum number of $\bar{n} = 0.58$ and a ground state
population of 63\%. Thus a situation at the crossover to a 2D atom
gas is reached.

Our experiment is presently limited by two main factors: First,
the unpolarized Cs sample in the $F=3$ ground state is not stable
against inelastic two-body collisions \cite{Guery-Odelin1998a},
which leads to heating and trap loss. Second, the
horizontal-vertical thermalization by elastic collisions becomes
inefficient for $\tau \lesssim 1$ \cite{Bouchoule2002a}. In our
horizontal evaporation scheme this inhibits vertical cooling when
two-dimensionality is appoached and may thus explain our present
limitation to $\tau \approx 1$.

In future experiments,  
both limitations can be overcome by polarizing the atoms into the
lowest substate ($F=3$, $m=3$). The polarized sample will be
stable against two-body decay and, moreover, offers magnetic
tunability of $s$-wave scattering by low-field Feshbach resonances
\cite{Kerman2001a}. Gravity can be compensated by using magnetic
gradients so that atoms can be evaporated vertically out of the
trap. With these modifications, quantum degeneracy of cesium may
be reached in the DEW trap under conditions where the trap
supports just a single vertical bound state. This would constitute
a unique system with strong analogies to H on liquid He, but with
full optical access and magnetically tunable interactions.

In conclusion, we have realized a cold and dense gas of Cs atoms
in a highly anisotropic optical trap with intriguing prospects for
experiments on 2D quantum gases. By evaporative cooling we have
already reached conditions at the crossover to two-dimensionality.
With further improvements the system opens up a new road to study
the fascinating and widely unexplored properties of degenerate
quantum gases in two dimensions.

\begin{acknowledgments}
We gratefully acknowledge support by the Austrian Science Fund
(FWF) within SFB 15 (project part 15).
\end{acknowledgments}


\begin{thebibliography}{27}
\expandafter\ifx\csname
natexlab\endcsname\relax\def\natexlab#1{#1}\fi
\expandafter\ifx\csname bibnamefont\endcsname\relax
  \def\bibnamefont#1{#1}\fi
\expandafter\ifx\csname bibfnamefont\endcsname\relax
  \def\bibfnamefont#1{#1}\fi
\expandafter\ifx\csname citenamefont\endcsname\relax
  \def\citenamefont#1{#1}\fi
\expandafter\ifx\csname url\endcsname\relax
  \def\url#1{\texttt{#1}}\fi
\expandafter\ifx\csname
urlprefix\endcsname\relax\def\urlprefix{URL }\fi
\providecommand{\bibinfo}[2]{#2}
\providecommand{\eprint}[2][]{\url{#2}}

\bibitem[{\citenamefont{Bagnato and Kleppner}(1991)}]{Bagnato1991a}
\bibinfo{author}{\bibfnamefont{V.}~\bibnamefont{Bagnato}} \bibnamefont{and}
  \bibinfo{author}{\bibfnamefont{D.}~\bibnamefont{Kleppner}},
  \bibinfo{journal}{Phys. Rev. A} \textbf{\bibinfo{volume}{44}},
  \bibinfo{pages}{7439} (\bibinfo{year}{1991}).

\bibitem[{\citenamefont{Kagan et~al.}(1987)\citenamefont{Kagan, Svistunov, and
  Shlyapnikov}}]{Kagan1987a}
\bibinfo{author}{\bibfnamefont{Y.}~\bibnamefont{Kagan}},
  \bibinfo{author}{\bibfnamefont{B.}~\bibnamefont{Svistunov}},
  \bibnamefont{and}
  \bibinfo{author}{\bibfnamefont{G.}~\bibnamefont{Shlyapnikov}},
  \bibinfo{journal}{Sov. Phys. JETP} \textbf{\bibinfo{volume}{66}},
  \bibinfo{pages}{314} (\bibinfo{year}{1987}).

\bibitem[{\citenamefont{Petrov et~al.}(2000)\citenamefont{Petrov, Holzmann, and
  Shlyapnikov}}]{Petrov2000a}
\bibinfo{author}{\bibfnamefont{D.}~\bibnamefont{Petrov}},
  \bibinfo{author}{\bibfnamefont{M.}~\bibnamefont{Holzmann}}, \bibnamefont{and}
  \bibinfo{author}{\bibfnamefont{G.}~\bibnamefont{Shlyapnikov}},
  \bibinfo{journal}{Phys. Rev. Lett.} \textbf{\bibinfo{volume}{84}},
  \bibinfo{pages}{2551} (\bibinfo{year}{2000}).

\bibitem[{\citenamefont{Walraven}()}]{Walraven1991a}
\bibinfo{author}{\bibfnamefont{J.~T.~M.} \bibnamefont{Walraven}},
  \bibinfo{howpublished}{in {\it Fundamental Systems in Quantum Optics}, edited
  by J. Dalibard, J.M. Raimond, and J. Zinn-Justin (Elsevier, Amsterdam, 1992),
  p. 485.}

\bibitem[{\citenamefont{Safonov et~al.}(1998)\citenamefont{Safonov, Vasilyev,
  Yasnikov, Lukashevich, and Jaakkola}}]{Safonov1998a}
\bibinfo{author}{\bibfnamefont{A.}~\bibnamefont{Safonov}}
\textit{et al.},
  \bibinfo{journal}{Phys. Rev. Lett.} \textbf{\bibinfo{volume}{81}},
  \bibinfo{pages}{4545} (\bibinfo{year}{1998}).

\bibitem[{\citenamefont{Vuleti{\'c} et~al.}(1998)\citenamefont{Vuleti{\'c},
  Chin, Kerman, and Chu}}]{Vuletic1998a}
\bibinfo{author}{\bibfnamefont{V.}~\bibnamefont{Vuleti{\'c}}},
  \bibinfo{author}{\bibfnamefont{C.}~\bibnamefont{Chin}},
  \bibinfo{author}{\bibfnamefont{A.~J.} \bibnamefont{Kerman}},
  \bibnamefont{and} \bibinfo{author}{\bibfnamefont{S.}~\bibnamefont{Chu}},
  \bibinfo{journal}{Phys. Rev. Lett.} \textbf{\bibinfo{volume}{81}},
  \bibinfo{pages}{5768} (\bibinfo{year}{1998}).

\bibitem[{\citenamefont{Bouchoule et~al.}(1999)\citenamefont{Bouchoule, Perrin,
  Kuhn, Morinaga, and Salomon}}]{Bouchoule1999a}
\bibinfo{author}{\bibfnamefont{I.}~\bibnamefont{Bouchoule}} \textit{et al.},
  \bibinfo{journal}{Phys. Rev. A} \textbf{\bibinfo{volume}{59}},
  \bibinfo{pages}{8(R)} (\bibinfo{year}{1999}).

\bibitem[{\citenamefont{Bouchoule et~al.}(2002)\citenamefont{Bouchoule,
  Morinaga, Salomon, and Petrov}}]{Bouchoule2002a}
\bibinfo{author}{\bibfnamefont{I.}~\bibnamefont{Bouchoule}},
  \bibinfo{author}{\bibfnamefont{M.}~\bibnamefont{Morinaga}},
  \bibinfo{author}{\bibfnamefont{C.}~\bibnamefont{Salomon}}, \bibnamefont{and}
  \bibinfo{author}{\bibfnamefont{D.}~\bibnamefont{Petrov}},
  \bibinfo{journal}{Phys. Rev. A} \textbf{\bibinfo{volume}{65}},
  \bibinfo{pages}{033402} (\bibinfo{year}{2002}).

\bibitem[{\citenamefont{G{\"orlitz} et~al.}(2001)\citenamefont{G{\"orlitz},
  Vogels, Leanhardt, Raman, Gustavson, Abo-Shaeer, Chikkatur, Gupta, Inouye,
  Rosenband et~al.}}]{Gorlitz2001a}
\bibinfo{author}{\bibfnamefont{A.}~\bibnamefont{G{\"orlitz}}}
\textit{et al.},
\bibinfo{journal}{Phys. Rev. Lett.}
  \textbf{\bibinfo{volume}{87}}, \bibinfo{pages}{130402}
  (\bibinfo{year}{2001}).

\bibitem[{\citenamefont{Gauck et~al.}(1998)\citenamefont{Gauck, Hartl,
  Schneble, Schnitzler, Pfau, and Mlynek}}]{Gauck1998a}
\bibinfo{author}{\bibfnamefont{H.}~\bibnamefont{Gauck}}
\textit{et al.},
  \bibinfo{journal}{Phys. Rev. Lett.} \textbf{\bibinfo{volume}{81}},
  \bibinfo{pages}{5298} (\bibinfo{year}{1998}).

\bibitem[{\citenamefont{Hammes et~al.}(2001)\citenamefont{Hammes, Rychtarik,
  and Grimm}}]{Hammes2001a}
\bibinfo{author}{\bibfnamefont{M.}~\bibnamefont{Hammes}},
  \bibinfo{author}{\bibfnamefont{D.}~\bibnamefont{Rychtarik}},
  \bibnamefont{and} \bibinfo{author}{\bibfnamefont{R.}~\bibnamefont{Grimm}},
  \bibinfo{journal}{C. R. Acad. Sci. Paris IV} \textbf{\bibinfo{volume}{2}},
  \bibinfo{pages}{625} (\bibinfo{year}{2001}).

\bibitem[{\citenamefont{Ovchinnikov et~al.}(1991)\citenamefont{Ovchinnikov,
  Shul'ga, and Balykin}}]{Ovchinnikov1991a}
\bibinfo{author}{\bibfnamefont{Y.~B.} \bibnamefont{Ovchinnikov}},
  \bibinfo{author}{\bibfnamefont{S.}~\bibnamefont{Shul'ga}}, \bibnamefont{and}
  \bibinfo{author}{\bibfnamefont{V.}~\bibnamefont{Balykin}},
  \bibinfo{journal}{J. Phys. B: At. Mol. Opt. Phys.}
  \textbf{\bibinfo{volume}{24}}, \bibinfo{pages}{3173} (\bibinfo{year}{1991}).

\bibitem[{\citenamefont{Desbiolles and Dalibard}(1996)}]{Desbiolles1996a}
\bibinfo{author}{\bibfnamefont{P.}~\bibnamefont{Desbiolles}} \bibnamefont{and}
  \bibinfo{author}{\bibfnamefont{J.}~\bibnamefont{Dalibard}},
  \bibinfo{journal}{Opt. Commun.} \textbf{\bibinfo{volume}{132}},
  \bibinfo{pages}{540} (\bibinfo{year}{1996}).

\bibitem[{\citenamefont{Perrin et~al.}()\citenamefont{Perrin, Colombe, Mercier,
  and Lorent}}]{Lorent2002a}
\bibinfo{author}{\bibfnamefont{H.}~\bibnamefont{Perrin}},
  \bibinfo{author}{\bibfnamefont{Y.}~\bibnamefont{Colombe}},
  \bibinfo{author}{\bibfnamefont{B.}~\bibnamefont{Mercier}}, \bibnamefont{and}
  \bibinfo{author}{\bibfnamefont{V.}~\bibnamefont{Lorent}},
  \bibinfo{howpublished}{Int.\ Workshop on Quantum Gases, Insel Reichenau, 19
  -21 July 2001, book of abstracts, p. 54}.

\bibitem[{\citenamefont{Mabuchi and Kimble}(1994)}]{Mabuchi1994a}
\bibinfo{author}{\bibfnamefont{H.}~\bibnamefont{Mabuchi}} \bibnamefont{and}
  \bibinfo{author}{\bibfnamefont{H.}~\bibnamefont{Kimble}},
  \bibinfo{journal}{Opt. Lett.} \textbf{\bibinfo{volume}{19}},
  \bibinfo{pages}{749} (\bibinfo{year}{1994}).

\bibitem[{\citenamefont{Domokos and Ritsch}(2001)}]{Domokos2001a}
\bibinfo{author}{\bibfnamefont{P.}~\bibnamefont{Domokos}} \bibnamefont{and}
  \bibinfo{author}{\bibfnamefont{H.}~\bibnamefont{Ritsch}},
  \bibinfo{journal}{Europhys. Lett.} \textbf{\bibinfo{volume}{54}},
  \bibinfo{pages}{306} (\bibinfo{year}{2001}).

\bibitem[{\citenamefont{C{\^o}t{\'e} and Segev}(2002)}]{cote2002a}
\bibinfo{author}{\bibfnamefont{R.}~\bibnamefont{C{\^o}t{\'e}}}
  \bibnamefont{and} \bibinfo{author}{\bibfnamefont{B.}~\bibnamefont{Segev}},
   \bibinfo{howpublished}{submitted for publication}.

\bibitem[{\citenamefont{Barnett et~al.}(2000)\citenamefont{Barnett, Smith,
  Olshanii, Johnson, Adams, and Prentiss}}]{Barnett2000b}
\bibinfo{author}{\bibfnamefont{A.~H.} \bibnamefont{Barnett}}
\textit{et al.},
  \bibinfo{journal}{Phys. Rev. A} \textbf{\bibinfo{volume}{61}},
  \bibinfo{pages}{023608} (\bibinfo{year}{2000}).

\bibitem[{\citenamefont{Burke et~al.}(2002)\citenamefont{Burke, Chu, Bryant,
  Williams, and Julienne}}]{Burke2002a}
\bibinfo{author}{\bibfnamefont{J.~P.} \bibnamefont{Burke}},
  \bibinfo{author}{\bibfnamefont{S.-T.} \bibnamefont{Chu}},
  \bibinfo{author}{\bibfnamefont{G.~W.} \bibnamefont{Bryant}},
  \bibinfo{author}{\bibfnamefont{C.~J.} \bibnamefont{Williams}},
  \bibnamefont{and} \bibinfo{author}{\bibfnamefont{P.~S.}
  \bibnamefont{Julienne}}, \bibinfo{journal}{Phys. Rev. A}
  \textbf{\bibinfo{volume}{65}}, \bibinfo{pages}{043411}
  (\bibinfo{year}{2002}).

\bibitem[{\citenamefont{Hammes et~al.}()\citenamefont{Hammes, Rychtarik,
  N{\"a}gerl, and Grimm}}]{Hammes2002a}
\bibinfo{author}{\bibfnamefont{M.}~\bibnamefont{Hammes}},
  \bibinfo{author}{\bibfnamefont{D.}~\bibnamefont{Rychtarik}},
  \bibinfo{author}{\bibfnamefont{H.-C.} \bibnamefont{N{\"a}gerl}},
  \bibnamefont{and} \bibinfo{author}{\bibfnamefont{R.}~\bibnamefont{Grimm}},
\eprint{physics/0204026}.

\bibitem[{\citenamefont{Landragin et~al.}(1996)\citenamefont{Landragin,
  Courtois, Labeyrie, Vansteenkiste, Westbrook, and Aspect}}]{Landragin1996a}
\bibinfo{author}{\bibfnamefont{A.}~\bibnamefont{Landragin}},
\textit{et al.},
  \bibinfo{journal}{Phys. Rev. Lett.} \textbf{\bibinfo{volume}{77}},
  \bibinfo{pages}{1464} (\bibinfo{year}{1996}).

\bibitem[{\citenamefont{Shimizu}(2001)}]{Shimizu2001a}
\bibinfo{author}{\bibfnamefont{F.}~\bibnamefont{Shimizu}},
  \bibinfo{journal}{Phys. Rev. Lett.} \textbf{\bibinfo{volume}{86}},
  \bibinfo{pages}{987} (\bibinfo{year}{2001}).

\bibitem[{\citenamefont{Ovchinnikov et~al.}(1997)\citenamefont{Ovchinnikov,
  Manek, and Grimm}}]{Ovchinnikov1997a}
\bibinfo{author}{\bibfnamefont{Y.~B.} \bibnamefont{Ovchinnikov}},
  \bibinfo{author}{\bibfnamefont{I.}~\bibnamefont{Manek}}, \bibnamefont{and}
  \bibinfo{author}{\bibfnamefont{R.}~\bibnamefont{Grimm}},
  \bibinfo{journal}{Phys. Rev. Lett.} \textbf{\bibinfo{volume}{79}},
  \bibinfo{pages}{2225} (\bibinfo{year}{1997}).

\bibitem[{\citenamefont{Friebel et~al.}(1998)\citenamefont{Friebel, DAndrea,
  Walz, Weitz, and H{\"a}nsch}}]{Friebel1998a}
\bibinfo{author}{\bibfnamefont{S.}~\bibnamefont{Friebel}}
\textit{et al.},
  \bibinfo{journal}{Phys. Rev. A} \textbf{\bibinfo{volume}{57}},
  \bibinfo{pages}{20(R)} (\bibinfo{year}{1998}).

\bibitem[{\citenamefont{Han et~al.}(2001)\citenamefont{Han, DePue, and
  Weiss}}]{Han2001a}
\bibinfo{author}{\bibfnamefont{D.}~\bibnamefont{Han}},
  \bibinfo{author}{\bibfnamefont{M.~T.} \bibnamefont{DePue}}, \bibnamefont{and}
  \bibinfo{author}{\bibfnamefont{D.}~\bibnamefont{Weiss}},
  \bibinfo{journal}{Phys. Rev. A} \textbf{\bibinfo{volume}{63}},
  \bibinfo{pages}{023405} (\bibinfo{year}{2001}); and D.~Weiss,
  private communication.

\bibitem[{\citenamefont{Gu{\'{e}}ry-Odelin
  et~al.}(1998)\citenamefont{Gu{\'{e}}ry-Odelin, S{\"{o}}ding, Desbiolles, and
  Dalibard}}]{Guery-Odelin1998a}
\bibinfo{author}{\bibfnamefont{D.}~\bibnamefont{Gu{\'{e}}ry-Odelin}},
  \bibinfo{author}{\bibfnamefont{J.}~\bibnamefont{S{\"{o}}ding}},
  \bibinfo{author}{\bibfnamefont{P.}~\bibnamefont{Desbiolles}},
  \bibnamefont{and} \bibinfo{author}{\bibfnamefont{J.}~\bibnamefont{Dalibard}},
  \bibinfo{journal}{Europhys. Lett.} \textbf{\bibinfo{volume}{44}},
  \bibinfo{pages}{26} (\bibinfo{year}{1998}).

\bibitem[{\citenamefont{Kerman et~al.}(2001)\citenamefont{Kerman, Chin,
  Vuleti{\'c}, Chu, Leo, Williams, and Julienne}}]{Kerman2001a}
\bibinfo{author}{\bibfnamefont{A.} \bibnamefont{Kerman}}
\textit{et al.},
\bibinfo{journal}{C. R. Acad. Sci. Paris IV}
  \textbf{\bibinfo{volume}{2}}, \bibinfo{pages}{633} (\bibinfo{year}{2001}).
\end{thebibliography}

\end{document}